\documentclass{amsart}
\usepackage{epsfig}
\usepackage{amsmath,amssymb}
\usepackage{slashed}
\usepackage{graphicx}
\numberwithin{equation}{section}

\usepackage{color}
\usepackage{hyperref}
\hypersetup{
 colorlinks=true,
 linkcolor=red,
 urlcolor=blue}

\usepackage{todonotes}

\newtheorem{theorem}{Theorem}[section]

\allowdisplaybreaks \numberwithin{equation}{section}

\newcommand{\PP}{{\mathbb P}}

\def\Jac{\mathop{\rm Jac}\nolimits}

\begin{document}

\title[Spectral Curves are Transcendental]{Spectral Curves are Transcendental}
\author{H.W. Braden}
\address{School of Mathematics, Edinburgh University, Edinburgh.}
\email{hwb@ed.ac.uk}

\begin{abstract}
Some arithmetic properties of spectral curves are discussed: the spectral curve, for  example,
of a charge $n\ge2$ Euclidean BPS monopole  is not defined over $\overline{\mathbb{Q}}$ if smooth.
\end{abstract}

\maketitle

\section{Introduction}

A fundamental ingredient of the modern theory of integrable systems is a curve, the \emph{spectral curve}, 
and the function theory of this curve enables (via the Baker-Akhiezer function, for example) the solution of
the system. Typically analytic properties of this curve are in the fore: here we will focus on a less well-developed aspect, its arithmetic properties. We will show that for an integrable system of interest the associated spectral curves are not defined over $\overline{\mathbb{Q}}$, the \emph{transcendental} of the title. This aspect is a manifestation of why it is so difficult to construct specific examples of some
systems. The result proven here depends on a number of deep results across several mathematical disciplines and what is novel is bringing them together. For a number theorist the transcendence of
periods is familiar: this paper provides a number of new examples where this is relevant. For an algebraic geometer, defining a curve by properties of lines bundles over it is not new: we see here the arithmetic consequences of this. To be concrete we will focus on a particular integrable system and remark on other examples. Neither a detailed knowledge of this particular physical system nor the arcane lore of integrable systems will be needed to understand this paper. 

The integrable system in focus here is that associated with Nahm's equations and BPS monopoles on 
$\mathbb{R}\sp3$, a reduction of the anti-self-dual Yang-Mills equations \cite{atiyah_hitchin_book}; for simplicity we will focus 
only on the case where the gauge group is $SU(2)$.   Some years ago Hitchin \cite{hitchin_83} gave a description of the regular solutions to this system in terms of a spectral curve $\mathcal{C}\subset T\mathbb{P}\sp1$  subject to constraints.  (These constraints will be reviewed later in the paper.)  Although the mathematics associated with these 
equations has proven remarkably rich, for example the moduli space of solutions may be given a 
hyperk\"ahler structure \cite{atiyah_hitchin_book}, the number of spectral curves that can be explicitly written down are few. Table 1
gives the list of those constructed over a period of some 35 years (see \cite{manton_sutcliffe_book}[Ch. 8] for references). Here $\eta$ and $\zeta$ are the fibre coordinate and affine base coordinate of $T\mathbb{P}\sp1$ and the degree of $\eta$ is the \lq\lq charge\rq\rq of the monopole. For these introductory comments let us focus on the charge $2$ BPS monopole and return to the others later in the text. Here we have a one parameter family of solutions  
\begin{equation*}
0=  \eta^2+\frac{\boldsymbol{K}(k)^2}4\left( \zeta^4+2(k^2-k'^2)\zeta^2+1\right),
\end{equation*}
where $K(k)$ is the complete elliptic integral with elliptic modulus $k$. 
The scalings of $\eta$ and $\zeta$ here are fixed by the constraints we have mentioned.  
With these normalisations
this curve is not expressible over $\overline{\mathbb{Q}}$: for if $k\not\in \overline{\mathbb{Q}}$
then at least one of  $k\, K(k)$  or $K(k)$ must be transcendental; finally a theorem of Schneider says that
if $k$ 
is algebraic, then  $K(k)$ is transcendental. We say the curve is transcendental.

Our goal is to establish
\begin{theorem}\label{bpsarithmetic}
Let  $\mathcal{C}$ be a smooth spectral curve of a charge $n\ge2$ Euclidean BPS monopole. Then  $\mathcal{C}$ is not defined over $\overline{\mathbb{Q}}$.
\end{theorem}
The theorem is a consequence of work of W\"ustholz on the vanishing or transcendence of certain periods,
and the work of a number of authors in developing Hitchin's constraints. We shall review this material next.
We remark that Hitchin's  construction of harmonic maps from the torus into the three sphere
also embodies transcendental constraints on a spectral curve \cite{hitchin_90}.

\begin{table}
\makebox[\textwidth]{
\begin{tabular}{|c| l |  l | c |}
\hline
&\hfil{Curve}\hfil& &\text{Symmetry}\\
\hline
1&$\displaystyle{0=\eta \prod_{l=1}\sp{m}(\eta^2+l^2\pi^2 \zeta^2)}$ &
&{Rotational}
\\
2&$\displaystyle{0= \prod_{l=0}\sp{m}(\eta^2+[l+\frac12]^2\pi^2 \zeta^2)}$  &
&\text{Rotational}
\\
&&&\\
3&$\displaystyle{0= \eta^2+\frac{\boldsymbol{K}(k)^2}4\left( \zeta^4+2(k^2-k'^2)\zeta^2+1\right) }$ &
&-\\
&&&\\
4&$\displaystyle{0=\eta^3+a_3 (\zeta^6 +5\sqrt{2}\zeta^3-1)}$
   &$\displaystyle{a_3=\pm \frac 1{48\sqrt{3} \pi^{3/2}} \, \Gamma\left(\frac16\right)^3\Gamma\left(\frac13\right)^3 }$   &Tetrahedral \\
  &&  $\displaystyle{\quad\,     =\pm \frac 1{48\sqrt{3}} \, B\left(\frac16,\frac13\right)^3 }$ &\\
 &&&\\
5&$\displaystyle{0=\eta^4+a_4 (\zeta^8+14 \zeta^4+1)}$  
  &$\displaystyle{a_4= \frac{3}{1024 \pi^2} \, \Gamma\left(\frac14\right)^8  
                                  =\frac{3}{256}\, B\left(\frac14,\frac12\right)^4  } $
   &Octahedral\\
&&&\\
6&$\displaystyle{0=\eta\left( \eta^4-4 a_4 (\zeta^8+14 \zeta^4+1)\right)}$  &&
Octahedral\\
 &&&\\
7&$\displaystyle{0=\eta\left( \eta^6+a_7\zeta (\zeta^{10}+11 \zeta^5-1)\right)}$
  &$\displaystyle{a_7= \frac 1{64\pi^{3}} \, \Gamma\left(\frac16\right)^6\Gamma\left(\frac13\right)^6 
                                    }$
 & Icosahedral\\
   &&  $\displaystyle{\quad\,        =\frac 1{64} \, B\left(\frac16,\frac13\right)^6 }$ &\\
&&&\\
8&$\displaystyle{0=\eta^4+36i a \kappa^3 \eta\zeta( \zeta^4-1)+3\kappa^4(\zeta^8+14 \zeta^4+1)}$ 
  &$\displaystyle{a\in\left(- {\sqrt{2}} / {3^{5/4}}, {\sqrt{2}} / {3^{5/4}}\right)}$, 
   &Tetrahedral $a\ne0$\hfill \\
&&\qquad $\kappa$\  \textrm{real half period of} 
&Octahedral $a=0$ \\
&&$\wp'(z)^2=4\wp(z)^3- 4\wp(z) +12 a^2$ &\\
&&& \\
9&$\displaystyle{0=\eta^3- 6(a^2+4\epsilon)^{1/3} \kappa^2 \eta\zeta^2 +2i\kappa^3 a(\zeta^5-\zeta)}$&
$a\in\mathbb{R},\  \epsilon=\pm1$, & ($x_1,x_2,x_3)\rightarrow$\\
&&\qquad $\kappa$\  \textrm{real half period of} &\quad $ (x_2, -x_1, -x_3)$\\
&&$\wp'(z)^2=4\wp(z)^3-3(a^2+4\epsilon)\sp{2/3}\wp(z) +4\epsilon $&
$a=\pm2, \epsilon=-1$, \\ & && Tetrahedral\\
&&& \\
10&$\displaystyle{0=\eta^3+\alpha \eta \zeta^2+\beta \zeta^6 +\gamma \zeta^3 -\beta}$& 
\textrm{With $(a,g)=(\alpha/\beta^{2/3}, \gamma/\beta)$
we obtain}
&$C_3$\\
&&\textrm{$g=g(a)$ by requiring a particular period}&$\alpha=0$, \textrm{Tetrahedral} \\
&&\textrm{vanish on the quotient genus 2 curve}& \\
&&& \\
\hline
\end{tabular}
}
\vskip 0.2in 
\caption{The known spectral curves}
\end{table}

\section{Arithmetic properties Curves and a theorem of W\"ustholz}

Here we briefly review several of the key ideas that lead to W\"ustholz's theorem on the transcendence of
periods in our setting of
a smooth algebraic curve $\mathcal{C}$: notably, that to $\mathcal{C}$ we may associate a commutative
algebraic group; to a commutative algebraic group defined over a number field W\"ustholz 
defines the notion of an analytic subgroup and gives necessary and sufficient conditions for the
existence of nontrivial algebraic subgroups; using this, and a theorem of Faltings-W\"ustholz,
the vanishing or transcendence of certain periods follows. W\"ustholz's theorems were outlined in \cite{wustholz_86} and a more detailed exposition may be found in \cite{baker_wustholz}.

First, to a smooth algebraic curve we may associate several commutative algebraic groups.
Following Rosenlicht we know that a commutative algebraic group $G$ may be expressed as the 
extension of an abelian variety $A$ by a linear algebraic group $L$ . Now $L$  is the product of a vector space and a torus, and so (following a possible base change) we may express $L=G_a\sp{r}\times G_m\sp{s}$ where $G_a$ is the additive group and $G_m$ the multiplicative group.  Then
$$0\longrightarrow G_a\sp{r}\times
G_m\sp{s}\longrightarrow G\longrightarrow
A\longrightarrow 0.$$
Now given a smooth algebraic curve $\mathcal{C}$ we have its associated Jacobian $\Jac(\mathcal{C})$,
a principally polarised abelian variety. The Jacobian may be described in a number of ways, 
one of which is in terms of differentials of the first kind, the regular differentials (over $\mathbb{C}$ these are just the usual holomorphic differentials). When one further considers 
differentials of the second kind, a differential that is the sum of an exact differential and one with vanishing
residues, one obtains a generalized Jacobian which is the extension of $\Jac(\mathcal{C})$ by the additive group $G_a$. (In characteristic zero the dimension of the vector space
of second kind differentials modulo exact differentials is twice the genus of $\mathcal{C}$, giving an upper bound on $r$ above in this context.) Finally
considering differentials of the third kind, those with simple poles, one gets an extension of $\Jac(\mathcal{C})$ by a multiplicative group $G_m$.
Generalized Jacobians also arise when we consider algebraic curves with singular points.
In our ensuing application to BPS monopoles we shall consider the commutative algebraic group
associated with holomorphic differentials. Faltings and W\"ustholz  \cite{faltings_wustholz_84} 
showed that if $X$  is a smooth quasi-projective variety over a number field
$\mathbb{K}$ possessing a $\mathbb{K}$-rational point and
 $\omega\in H\sp0(X,\Omega\sp1_{X/\mathbb{K}})$ a closed holomorphic differential on $X$
then $\omega$ is the ÒpullbackÓ of an invariant differential form on a generalized  Albanese variety.
That is Faltings and W\"ustholz provide a map $(X,\omega)$ to a commutative algebraic group over $\mathbb{K}$; we are just specialising to the case when $X$ is a curve.

Now let $G$ be a commutative algebraic group defined over a number field
$\mathbb{K}$ with Lie algebra $\mathfrak{g}=$Lie($G$). If $\mathfrak{b}$ is a subalgebra of
$\mathfrak{g}$, set
\begin{equation*} B:=\exp_G(\mathfrak{b}\otimes_\mathbb{K}\mathbb{C})\le
G(\mathbb{C}). \end{equation*}
Following W\"ustholz we say that $B$ is an \emph{analytic subgroup}  of $G$ defined over $\mathbb{K}$; it is not necessarily a
closed subgroup of $G(\mathbb{C})$. One wishes to determine the group of algebraic points
$$B(\overline{\mathbb{K}}):=B\cap G(\overline{\mathbb{K}}).$$
Certainly this group is nontrivial if there exists a nontrivial algebraic subgroup $0\ne H\le G$ defined over a number field such that $H(\mathbb{C})\leq B$,  for then $$0\ne H(\overline{\mathbb{K}})\leq B(\overline{\mathbb{K}}).$$
The analytic subgroup theorem provides the converse of this.
\begin{theorem}[Analytic Subgroup Theorem (W\"ustholz)] Let
$B\subseteq G(\mathbb{C})$ be an analytic subgroup defined over
$\mathbb{K}$. Then $ B(\overline{\mathbb{K}})\ne0$ if and only if there
exists a nontrivial algebraic subgroup $H\leq G$ defined over a
number field such that $H(\mathbb{C})\leq B$.
\end{theorem}
Using the analytic subgroup theorem and the Faltings-W\"ustholz result   W\"ustholz  deduces that

 \begin{theorem}[W\"ustholz \cite{wustholz_86}] \label{wustholzperiods}Let $(X,\omega)$ be as previously described. Then $\int_\gamma \omega$ \ ($\gamma\in
H_1(X,\mathbb{Z})$) are either zero or transcendental.
\end{theorem}

This theorem yields many of the classical transcendence results (see \cite{baker_wustholz}[\S6.3])
including the theorem of Schneider, noted in the introduction, that the periods of an elliptic integral
with rational elliptic modulus is transcendental. We will apply the theorem by transforming one of Hitchin's
constraints into a constraint on holomorphic differentials.

\section{Monopoles and Hitchin's Constraints}

As already noted, BPS magnetic monopoles describe a class of finite energy solutions to a reduction of the anti-self-dual Yang-Mills equations \cite{atiyah_hitchin_book,hitchin_83}. Assuming a static solution (where the connection is independent of the
\lq time\rq\ coordinate) these partial differential equations take the form
 $$\star F = D\Phi,$$
  where $F$ is the curvature of the connection $A$ for gauge group $G$ with Lie algebra $\mathfrak{g}$, $\Phi$ is a Higgs field, $\star$ is the Hodge-$\star$ operator for $\mathbb{R}^3$ (though other 3-manifolds may also be considered). Suitable boundary conditions need to be specified so as to ensure finiteness of the
energy;  these boundary conditions allow one to define the Higgs field over the 2-sphere 
\lq \lq at infinity\rq\rq\
and the  \lq\lq charge\rq\rq\ of the monopole is the first Chern class of this bundle. Two approaches exist to
the problem of constructing these solutions. Just as the self-duality equations may be understood in terms
of twistor theory, a reduction of this exists describing monopoles, where mini-twistor space $ T\mathbb{P}\sp1$, the space of lines in $\mathbb{R}^3$, plays the corresponding role. The zero-curvature equation
arising from the anti-self-dual Yang-Mills equations leads to $[D_3-i\Phi, D_{\bar z}]=0$ and considering the
operator $D_3-i\Phi$ (which depends holomorphically on $z$). The collection of lines in $\mathbb{R}^3$ for which this operator has square integrable solutions forms a curve $\mathcal{C} \subset  T\mathbb{P}\sp1 $.
A second approach was discovered by Nahm in which the solutions to the partial differential equations were constructed in terms of solutions to a set of matrix ODE's (\lq\lq Nahm's Equations\rq\rq) and an associated (ordinary) differential operator built from these; this is the Nahm correspondence. Nahm's equations may be viewed as an integrable system and have a Lax pair formulation and corresponding spectral curve.  This
spectral curve is precisely the curve $\mathcal{C}$ arising from the mini-twistor viewpoint. Constructing regular solutions from both approaches becomes one of specifying $\mathcal{C}$ and it was
Hitchin \cite{hitchin_83} who gave necessary and sufficient algebro-geometric constraints on the spectral curve of this integrable system to yield BPS monopoles. 

In terms of the coordinates $ (\eta,\zeta)\rightarrow \eta\frac{d}{d\zeta}\in T\mathbb{P}\sp1 $ the
spectral curve $\mathcal{C}$ is specified by the vanishing of the polynomial $P(\eta,\zeta)$ where
$$P(\eta,\zeta)=\eta\sp{n}+a_1(\zeta)\eta\sp{n-1}+\ldots+a_n(\zeta),\qquad
\deg a_r(\zeta)\le2r.$$
This curve, which we will assume smooth, has genus $({n}-1)\sp2$. We note that $T\mathbb{P}\sp1 $
has the antiholomorphic involution  $\iota:(\eta,\zeta)\rightarrow(-\bar\eta/{\bar\zeta}\sp2,-1/{\bar\zeta})$
which reverses the orientation of lines. We may cover $\pi:T\mathbb{P}\sp1 \rightarrow \mathbb{P}\sp1$ by the two patches $\widehat{\mathcal{U}}_{0,1}$ corresponding to the pre-images of the standard cover  $\mathcal{U}_{0,1}$ of  $\mathbb{P}\sp1$. Let $\mathcal{L}^{\lambda}(m)$ the holomorphic line
bundle on $T\PP\sp1$ with transition function $g_{01}=\zeta^m\exp{(-\lambda\eta/\zeta)}$; setting
$\mathcal{L}^{\lambda}:= \mathcal{L}^{\lambda}(0)$, then
$\mathcal{L}^{\lambda}(m)\equiv\mathcal{L}^{\lambda}\otimes\pi\sp*\mathcal{O}(m)$.
Hitchin's constraints are then:

\begin{description}
\item[H1] $\mathcal{C}$ is real with respect to $\iota$,

\item[H2] {$\mathcal{L}^2$ is trivial on $\mathcal{C}$ and $\mathcal{L}\sp1(n-1)$ is real,}

\item[H3] {$H^0(\mathcal{C},\mathcal{L}^{s}(n-2))=0$}
    for $s\in(0,2)$.
\end{description}
Here the parameter $s$ describing the linear flow of Hitchin's line bundles corresponds to the \lq time\rq\
 of the integrable systems evolution, this linear evolution being described by a straight line in $\Jac(\mathcal{C})$. The third condition says that this \emph{real} straight line does not intersect the
 theta divisor for $s\in(0,2)$, while it does at $s=0,2$. Only the first of these constraints is easily
  implemented. The reality conditions {\bf H1} mean
${a_r(\zeta)=(-1)\sp{r}\zeta\sp{2r}\overline{a_r(-\frac1{\bar\zeta})}}$
and as a consequence $a_r(\zeta)$ is given by $2r+1$ (real) parameters. It is the difficulty of 
making effective {\bf H2,\,3} that makes the construction of monopoles so difficult.

Ercolani and Sinha \cite{ercolani_sinha_89} made the initial study of {\bf H2}. 
The triviality of $\mathcal{L}^2$  means that
there exists a nowhere-vanishing holomorphic section; in terms of our cover and transition functions we have 
$f_{0}(\eta,\zeta)=\mathrm{exp} \left\{-2 {\eta}/{\zeta}\right\} f_1(\eta,\zeta)$ with $f_i$ holomorphic in
$\widehat{\mathcal{U}}_i$. The logarithmic differential of $f_0$ thus yields a meromorphic differential
for which $\mathrm{exp}\oint\limits_{\gamma}\mathrm{d}\mathrm{log}\,f_{0}=1$ for all $\gamma\in
H_1(\mathbb{Z},\mathcal{C})$, and the flow in the Jacobian is governed by the meromorphic differential
$$ \gamma_\infty(P)=\frac{1}{2}\,\mathrm{d}\mathrm{log} \, f_{0}(P)+\imath \pi\,\sum_{j=1}^g m_j\,\omega_j(P).$$
Here the $\omega_i$ are canonically $\mathfrak{a}$-normalized holomorphic
differentials ($\oint_{\mathfrak{a}_k}\omega_j=\delta_{jk} $)  and we  add an appropriate linear combination
so that $\oint\limits_{\mathfrak{a}_k}\gamma_\infty=0$. These observations, together with the Riemann
bilinear relations yield
\begin{theorem}[Ercolani-Sinha Constraints 
\cite{ercolani_sinha_89, houghton_manton_romao_00,Braden2010c}]\label{ES}
The following are equivalent:
\begin{enumerate}
 \item$\mathcal{L}\sp2$ is trivial on $\mathcal{C}$.

\item$2\boldsymbol{U}\in \Lambda\Longleftrightarrow$ $
    \boldsymbol{U}=\frac{1}{2\pi\imath}\left(\oint_{\mathfrak{b}_1}\gamma_{\infty},
    \ldots,\oint_{\mathfrak{b}_g}\gamma_{\infty}\right)\sp{T}=
    \frac12 \boldsymbol{n}+\frac12\tau\boldsymbol{m} , $ where $\Lambda$ is the period lattice.

\item
There exists a 1-cycle $\mathfrak{es}=\boldsymbol{n}\cdot{\mathfrak{a}}+
    \boldsymbol{m}\cdot{\mathfrak{b}}$ such that  every  holomorphic differential
    $$\Omega=\dfrac{\beta_0\eta^{n-2}+\beta_1(\zeta)\eta^{n -3}+\ldots+\beta_{n-2}(\zeta)}{\frac{\partial
    {P}}{\partial \eta}}\,d\zeta$$
    has period $\oint\limits_{\mathfrak{es}}\Omega=-2\beta_0 $. This 1-cycle satisfies
    $\iota_\ast \mathfrak{es}=-\mathfrak{es}$.
    
\end{enumerate}
\end{theorem}
We may now prove theorem (\ref{bpsarithmetic}). Suppose $\mathcal{C}$, and so the polynomial $P(\eta,\zeta)$, is defined over $\overline{\mathbb{Q}}$. We may let $\mathbb{K}$ be the a number field that
contains the coefficients of $P$ and the roots of $P(0,\zeta)=a_n(\zeta)$; thus $\mathcal{C}$ contains
a $\mathbb{K}$-rational point. Consider the holomorphic differential $\omega=\left({\eta^{n-2}}/{\frac{\partial{P}}{\partial\eta}}\right)\,d\zeta$ (recall $n\ge2$ in the theorem). 
We are assuming $\mathcal{C}$ smooth and so the conditions of theorem
(\ref{wustholzperiods}) are satisfied, thus the periods of $\omega$ are either zero or transcendental.
But this contradicts theorem (\ref{ES}) and so  $\mathcal{C}$ cannot be defined over 
$\overline{\mathbb{Q}}$.  We say the Ercolani-Sinha constraints impose $g$ \emph{transcendental constraints} on the curve.

A number of remarks are perhaps in order.
\begin{enumerate}

\item Hitchin's constraints do not require $\mathcal{C}$ to be irreducible and a number of the examples of
Table 1 are in fact reducible.  These examples show that $\mathcal{C}$ is not defined over $\overline{\mathbb{Q}}$ here as well.

\item One can say more about $2\boldsymbol{U}$: it is in fact a primitive vector in the period lattice. By
tensoring with a section of  $\pi\sp*\mathcal{O}(n-2)\vert_\mathcal{C}$ we obtain a map
$\mathcal{O}(\mathcal{L}\sp{s})\hookrightarrow \mathcal{O}(\mathcal{L}\sp{s}(n-2))$ and so the
vanishing of $H\sp0\left(\mathcal{C},\mathcal{O}(\mathcal{L}^{s}(n-2))\right)$ also entails that
$ H\sp0\left(\mathcal{C},\mathcal{O}(\mathcal{L}\sp{s})\right)=0$ for $s\in(0,2)$; this means that
$2\boldsymbol{U}$ is in a primitive vector.

\item If $\mathcal{A}$ (respectively $\mathcal{B}$) denotes the matrix of  $\mathfrak{a}$- periods
(respectively $\mathfrak{b}$-periods) for a basis of holomorphic differentials this may be chosen
so that (with $\omega$ the final basis element)
$$(\mathbf{n},\mathbf{m})\begin{pmatrix}{\mathcal{A}}\\
{\mathcal{B}}\end{pmatrix}=-2(0,\ldots,0,1).$$
That is the Ercolani-Sinha constraints reflect rational relations between the periods.

\item  It is possible for a curve to satisfy {\bf H2} and yet fail  {\bf H3}; we shall give examples of this below.

\end{enumerate}

\section{Examples}

\subsection{The examples} The known spectral curves in Table 1 all exhibit symmetries; these may
simplify the problem. Reference \cite{Braden2011} shows how questions about the Ercolani-Sinha vector reduce to questions for the
quotient curve; the flows of the integrable system are also shown there to simplify using a theorem of Fay and Accola.
Examples 4-9 of Table 1 all exhibit a Platonic symmetry group \cite{hitchin_manton_murray_95}, which evidences itself in the Klein polynomials of the appropriate spectral curves; these curves all quotient to an elliptic curve.
The elliptic curves for the discrete monopole configurations of examples 4-8 each yield a Beta function of rational arguments,
the transcendence of which is also a result Schneider. The transcendence of the one-parameter families 8, 9
both follow by a similar argument to that of the introduction using Scheider's result on the transcendence of
the periods of the Weierstrass $\wp$-function for algebraic $g_{2,3}$.
Although the examples 1, 2 (for $n\ge3$), 6, 7 are for reducible curves and so outwith the theorem, they too are transcendental. The final curve has $C_3$ symmetry and
quotients over a genus 2 curve \cite{Braden2011a}. The transcendence of the periods here requires theorem (\ref{wustholzperiods}); a genus 2-variant of the AGM due to Richelot may be used for their computation.

\subsection{Solving {\bf H2} and number theory}
Here we shall give a countable number of examples that satisfy {\bf H2} but fail {\bf H3}  using some
wonderful results of Ramanujan; there are various open questions here. 

The trigonal family of genus 4 curves
$w^3 =\prod_{i=1}\sp6(z-z_i)$ has been studied by a number of
authors \cite{Picard1883,Wellstein1899,shiga_88,matsumoto_01} in connection with the
configuration of six points on $\mathbb{P}\sp1$. The symmetry $(z,w)\rightarrow(z, \rho w)$ 
($\rho=e\sp{2i\pi/3}$)
of this family allows the period matrix to be calculated in terms of four  parameters 
$x_i=\oint_{\mathfrak{a}_i}dz/w$,
the four $\mathfrak{a}$-periods of the differential $dz/w$,  for an appropriately chosen homology basis.
The Ercolani-Sinha constraints may \cite{Braden2010c} then be expressed as constraints on these $x_i$ 
of the form 
$$\frac{x_1}{n_1+\rho^2 m_1}=\frac{x_2}{n_2+\rho^2 m_2}=\frac{x_3}{n_3+\rho^2 m_3}=
\frac{x_4}{-n_4+\rho^2 m_4}.$$
These equations should be viewed as constraining the locus of the six points $z_i$; the Ercolani-Sinha constraints express algebraic dependence between these constraints and the difficult question is one of
realising these.  Of course this general curve need not satisfy  {\bf H1}. To make progress we do this
and consider the restricted family of curves
\begin{equation}\label{symmetric}
\eta^3+\chi(\zeta^6+b\zeta^3-1)=0,\qquad b,\chi\in\mathbb{R}.
\end{equation}
These curves have an additional $C_3$ symmetry. A consequence of this symmetry is that the previously
four independent periods $x_i$ are reduced to two (reflecting a genus 2 quotient curve) and that the
periods may be expressed in terms of hypergeometric functions. In this simpler setting we have

\begin{theorem}[Braden-Enolski \cite{Braden2010a}]
To each pair of relatively prime integers $(n,m)=1$ for which
$(m + n)(m -2n)<0$
we obtain a solution to the Ercolani-Sinha constraints for the 
curve (\ref{symmetric}) as follows. First we solve for $t$, where
\begin{equation}\label{esct}
\dfrac{2n-m}{m + n}=\frac{{_2F_1}(\frac{1}{3},
\frac{2}{3}; 1,t)}{{_2F_1}(\frac{1}{3}, \frac{2}{3}; 1,1-t)}.
\end{equation}
Then
$\displaystyle{
b=\frac{1-2t}{\sqrt{t(1-t)}}}$, $\displaystyle{t=
\frac{-b+\sqrt{b^2+4}}{2\sqrt{b^2+4}}.
}$
With $\alpha\sp6=t/(1-t)$ then
$$
\chi^{\frac{1}{3}} = -(n + m )\, \frac{2 \pi}{3
    \sqrt{3}}\ \frac{\alpha}{(1+\alpha\sp6)\sp\frac13}\ {_2F_1}(\frac{1}{3}, \frac{2}{3}; 1,
    t).
$$
\end{theorem}
At this stage we have a countable number of curves satisfying {\bf H1,2} provided we can solve
(\ref{esct}). Now towards proving (amongst others) the following formulae of Ramanujan,
$$
\frac{4}{\pi}=\sum_{m=0}\sp\infty \frac{(1+6
m)(\frac12)_m(\frac12)_m(\frac12)_m} {(m!)\sp3 4^m},\ \ \
{ \frac{27}{4\pi}=\sum_{m=0}\sp\infty \frac{(2+15
m)(\frac12)_m(\frac13)_m(\frac23)_m}
{(m!)\sp3\,\left(\frac{27}{2}\right)\sp{m}}} ,
$$
Bernd, Bhargava and Garvan
 \cite{berndt_bhargava_garvan_95} 
introduced the following extension of a modular equation of degree $n$: a
\emph{modular equation of degree $n$ and signature} $r$ ($r=2,3,4,6$)
 is defined to be a relation between $\alpha$, $\beta$ of the form
$$n\,\frac{_2F_1(\frac{1}{r},\frac{r-1}{r};1;1-\alpha)}
{_2F_1(\frac{1}{r},\frac{r-1}{r};1;\alpha)}=
\frac{_2F_1(\frac{1}{r},\frac{r-1}{r};1;1-\beta)}
{_2F_1(\frac{1}{r},\frac{r-1}{r};1;\beta)}.$$
For small prime $n$ they solve this; for example
\begin{align*}n=2,\ &r=3\Longrightarrow
(\alpha\beta)\sp{1/3}+((1-\alpha)(1-\beta))\sp{1/3}=1,
\end{align*}
whence for $\alpha=1/2$ then $\beta=\frac{1}{2}+\frac{5\sqrt{3}}{18}$. 
Using this theory we may solve (\ref{esct}); several small values are given in
Table 2.
\begin{table}\label{essymmetriccurve}
\caption{Values of $b$, $t$ solving the Ercloani-Sinha constraints for the curve (\ref{symmetric}).}
$$
\begin{array}{|c|c|c|c|c|} \hline
n&m&\frac{2n-m}{m+n}
&t&b\\
\hline 2&1&1&\frac{1}{2}&0\\
\hline
1&0&{2}&\frac{1}{2}+\frac{5\sqrt{3}}{18}&5\sqrt{2}\\
\hline1&1&\frac{1}{2}&\frac{1}{2}-\frac{5\sqrt{3}}{18}&5\sqrt{2}\\
 \hline 4&-1&3&(63+171\sqrt[3]{2}-18\sqrt[3]{4})/250&
\frac13(44+38\sqrt[3]{2}+26\sqrt[3]{4})\\
\hline 5&-2&4&\frac{1}{2}+\frac{153\sqrt{3}-99\sqrt{2}}{250}&
9\sqrt{458+187\sqrt{6}}\\
\hline
\end{array}
$$
\end{table}
We observe that for $|n|=1$ the curve has in fact tetrahedral symmetry.
The values of $b$ here will be algebraic so from Theorem (\ref{bpsarithmetic}) we have that $\chi$
must be transcendental, This is indeed a known result: it follows from the 
transcendence (for $t\in \mathbb{Q}$) of 
 $$\pi\times{_2F_1}\left(\frac13,\frac23;1;t\right)=
 \frac{\sqrt{3}}2 \int_0\sp1 u^{-1/3}(1-u)^{-2/3}(1-ut)^{-1/3}du,
 $$ 
 which again follows from W\"ustholz's theorem.
 In \cite{Braden2010b} an expression for $b$ is given in terms of Jacobi theta functions. We have 
 not discussed the final of Hitchin's constraints {\bf H3}. Apart from the the case of tetrahedral symmetry
 it is believed no member of this family satisfies {\bf H3} and a conjecture exists to the points
 $s$ for which the family of line bundles has sections.

\def\cprime{$'$} \def\cdprime{$''$} \def\cprime{$'$}


\begin{thebibliography}{10}

\bibitem{atiyah_hitchin_book}
Michael Atiyah and Nigel Hitchin.
\newblock {\em The geometry and dynamics of magnetic monopoles}.
\newblock M. B. Porter Lectures. Princeton University Press, Princeton, NJ,
  1988.

\bibitem{baker_wustholz}
A.~Baker and G.~W{\"u}stholz.
\newblock {\em Logarithmic forms and {D}iophantine geometry}, volume~9 of {\em
  New Mathematical Monographs}.
\newblock Cambridge University Press, Cambridge, 2007.

\bibitem{berndt_bhargava_garvan_95}
Bruce~C. Berndt, S.~Bhargava, and Frank~G. Garvan.
\newblock Ramanujan's theories of elliptic functions to alternative bases.
\newblock {\em Trans. Amer. Math. Soc.}, 347(11):4163--4244, 1995.

\bibitem{Braden2011}
H.~W. Braden.
\newblock Cyclic monopoles, affine {T}oda and spectral curves.
\newblock {\em Comm. Math. Phys.}, 308(2):303--323, 2011.

\bibitem{Braden2011a}
H.~W. Braden, Antonella D'Avanzo, and V.~Z. Enolski.
\newblock On charge-3 cyclic monopoles.
\newblock {\em Nonlinearity}, 24(3):643--675, 2011.

\bibitem{Braden2010b}
H.~W. Braden and V.~Z. Enolski.
\newblock On the tetrahedrally symmetric monopole.
\newblock {\em Comm. Math. Phys.}, 299(1):255--282, 2010.

\bibitem{Braden2010c}
H.~W. Braden and V.~Z. {\`E}nol{\cprime}ski{\u\i}.
\newblock {$\rm SU(2)$}-monopoles, curves with symmetries, and {R}amanujan's
  legacy.
\newblock {\em Mat. Sb.}, 201(6):19--74, 2010.

\bibitem{Braden2010a}
H.~W. Braden and T.~P. Northover.
\newblock Klein's curve.
\newblock {\em J. Phys. A}, 43(43):434009, 17, 2010.

\bibitem{ercolani_sinha_89}
N.~Ercolani and A.~Sinha.
\newblock Monopoles and {B}aker functions.
\newblock {\em Comm. Math. Phys.}, 125(3):385--416, 1989.

\bibitem{faltings_wustholz_84}
G.~Faltings and G.~W{\"u}stholz.
\newblock Einbettungen kommutativer algebraischer {G}ruppen und einige ihrer
  {E}igenschaften.
\newblock {\em J. Reine Angew. Math.}, 354:175--205, 1984.

\bibitem{hitchin_83}
N.~J. Hitchin.
\newblock On the construction of monopoles.
\newblock {\em Comm. Math. Phys.}, 89(2):145--190, 1983.

\bibitem{hitchin_90}
N.~J. Hitchin.
\newblock Harmonic maps from a {$2$}-torus to the {$3$}-sphere.
\newblock {\em J. Differential Geom.}, 31(3):627--710, 1990.

\bibitem{hitchin_manton_murray_95}
N.~J. Hitchin, N.~S. Manton, and M.~K. Murray.
\newblock Symmetric monopoles.
\newblock {\em Nonlinearity}, 8(5):661--692, 1995.

\bibitem{houghton_manton_romao_00}
C.~J. Houghton, N.~S. Manton, and N.~M. Rom{\~a}o.
\newblock On the constraints defining {BPS} monopoles.
\newblock {\em Comm. Math. Phys.}, 212(1):219--243, 2000.

\bibitem{manton_sutcliffe_book}
Nicholas Manton and Paul Sutcliffe.
\newblock {\em Topological solitons}.
\newblock Cambridge Monographs on Mathematical Physics. Cambridge University
  Press, Cambridge, 2004.

\bibitem{matsumoto_01}
Keiji Matsumoto.
\newblock Theta constants associated with the cyclic triple coverings of the
  complex projective line branching at six points.
\newblock {\em Publ. Res. Inst. Math. Sci.}, 37(3):419--440, 2001.

\bibitem{Picard1883}
Emile Picard.
\newblock Sur des fonctions de deux variables ind{\'e}pendantes analogues aux
  fonctions modulaires.
\newblock {\em Acta Mathematica}, 2(1):114--135, 1883.

\bibitem{shiga_88}
Hironori Shiga.
\newblock On the representation of the {P}icard modular function by {$\theta$}
  constants. {I}, {II}.
\newblock {\em Publ. Res. Inst. Math. Sci.}, 24(3):311--360, 1988.

\bibitem{Wellstein1899}
J.~Wellstein.
\newblock Zur theorie der functionenclasses3 = (s-$\alpha$2)...(z-$\alpha$6).
\newblock {\em Mathematische Annalen}, 52(2):440--448, 1899.

\bibitem{wustholz_86}
G.~W{\"u}stholz.
\newblock Algebraic groups, {H}odge theory, and transcendence.
\newblock In {\em Proceedings of the {I}nternational {C}ongress of
  {M}athematicians, {V}ol. 1, 2 ({B}erkeley, {C}alif., 1986)}, pages 476--483.
  Amer. Math. Soc., Providence, RI, 1987.

\end{thebibliography}
\end{document}